\documentclass[sigconf]{acmart}
\usepackage{graphicx} 
\usepackage{booktabs}
\usepackage{subcaption}
\usepackage{algorithm}
\usepackage{algpseudocode}
\usepackage{amsmath}
\usepackage{tikz}
\usetikzlibrary{positioning,shapes,arrows}
\usepackage{listings}
\title[Unifrying Ranking and Generation in Query Auto-Completion via RAG and Multi-Objective Alignment]{Unifying Ranking and Generation in Query Auto-Completion via Retrieval-Augmented Generation and Multi-Objective Alignment}

\author{Kai Yuan$^{1}$, Anthony Zheng$^{1}$, Jia Hu$^{1}$, Divyanshu Sheth$^{1}$, Hemanth Velaga$^{1}$, Kylee Kim$^{1}$, Matteo Guarrera$^{2}$, Besim Avci$^{1}$, Xuetao Yin$^{1}$, Jianhua Li$^{1}$, Rajyashree Mukherjee$^{1}$, Sean Suchter$^{1}$}
\affiliation{%
  \institution{\normalsize $^{1}$Apple \quad $^{2}$UC Berkeley}
  \country{USA}
}

\renewcommand{\shortauthors}{Yuan, Zheng, Hu, et al.}

\date{August 2025}

\acmConference[Preprint]{Preprint}{2026}{}
\acmBooktitle{Preprint}

\begin{document}

\begin{abstract}
Query Auto-Completion (QAC) suggests query completions as users type, helping them articulate intent and reach results more efficiently. Existing approaches face fundamental challenges: traditional retrieve-and-rank pipelines have limited long-tail coverage and require extensive feature engineering, while recent generative methods suffer from hallucination and safety risks. We present a unified framework that reformulates QAC as end-to-end list generation through Retrieval-Augmented Generation (RAG) and multi-objective Direct Preference Optimization (DPO). Our approach combines three key innovations: (1) reformulating QAC as end-to-end list generation with multi-objective optimization; (2) defining and deploying a suite of rule-based, model-based, and LLM-as-judge verifiers for QAC, and using them in a comprehensive methodology that combines RAG, multi-objective DPO, and iterative critique-revision for high-quality synthetic data; (3) a hybrid serving architecture enabling efficient production deployment under strict latency constraints. Evaluation on a large-scale commercial search platform demonstrates substantial improvements: offline metrics show gains across all dimensions, human evaluation yields +0.40 to +0.69 preference scores, and a controlled online experiment achieves 5.44\% reduction in keystrokes and 3.46\% increase in suggestion adoption, validating that unified generation with RAG and multi-objective alignment provides an effective solution for production QAC. This work represents a paradigm shift to end-to-end generation powered by large language models, RAG, and multi-objective alignment, establishing a production-validated framework that can benefit the broader search and recommendation industry.

\end{abstract}
\settopmatter{printacmref=false}
\renewcommand\footnotetextcopyrightpermission[1]{}

\maketitle

\section{Introduction}
\label{intro}
Query Auto-Completion (QAC) suggests query completions as users type, reducing keystrokes and helping them articulate intent more efficiently. In large-scale search systems it is often the first touchpoint of the search experience. Traditionally, QAC follows a two-stage retrieve-and-rank paradigm \cite{BarYossef2011, conf/sigir/Zhang2015, kai2021}: candidate generation extracts potential completions from historical query logs or product catalogs, followed by a learning-to-rank model \cite{Santu2017} that scores and selects the top-k suggestions based on relevance and popularity signals.

\begin{figure}[t]
    \centering
    \vspace*{0.1cm}
    \includegraphics[width=0.99\linewidth]{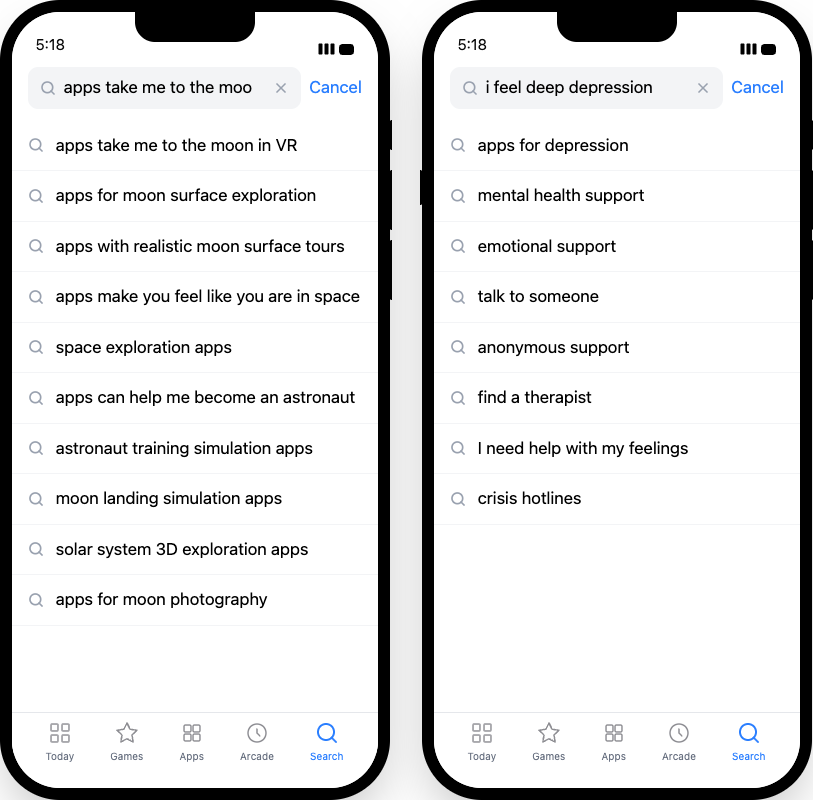}
    \caption{Illustrative QAC results in a mobile application search system. \textbf{Left:} for the prefix ``apps take me to the moo'', our generative QAC approach infers the user's intent to explore moon-related experiences and surfaces grounded alternatives such as VR simulations and space exploration apps, rather than hallucinating impossible ``go to the moon'' apps. \textbf{Right:} for the sensitive prefix ``i feel deep depression'', the system produces safe, supportive suggestions that encourage help-seeking (e.g., mental health support, crisis hotlines) and avoid harmful content.}
    \label{fig:intro}
\end{figure}
\vspace{\baselineskip}

Figure~\ref{fig:intro} highlights two representative cases from a mobile app search system. In the left example, the user types ``apps take me to the moo''; a naive generative model might hallucinate non-existent apps that literally take users to the moon, while a log-based system may have no suitable candidates. A high-quality QAC system should instead infer the underlying intent and propose helpful but catalog-grounded alternatives. In the right example, the user types ``i feel deep depression''; here the system must recognize a sensitive situation and surface safe, supportive suggestions rather than potentially harmful completions. These scenarios expose the limitations of conventional retrieve-and-rank pipelines and unconstrained generation in handling intent understanding, catalog groundedness, and safety simultaneously.

This conventional retrieve-and-rank approach, while effective for popular ``head'' queries, faces three fundamental limitations. \textbf{First}, its coverage is bounded by historical logs, making it incapable of serving high-quality suggestions for novel or long-tail prefixes with sparse interaction history \cite{Hasan2011}. \textbf{Second}, optimizing multiple competing objectives—relevance, diversity, safety, and groundedness—requires extensive manual feature engineering and heuristic blending strategies that fail to capture complex interdependencies. Among these objectives, \textit{groundedness} is particularly critical for closed-domain search scenarios such as mobile application search and music catalog search, where users rely on QAC to discover what actually exists in fixed catalogs. Suggestions that lead to empty results fundamentally break user trust. \textbf{Third}, the decoupled candidate generation and ranking stages prevent holistic list-level optimization, often resulting in redundant or semantically similar suggestions (e.g., ``workout apps for women'' vs. ``women workout apps'').

Recent work has explored generative language models \cite{Park2017, Sordoni2015, conf/kdd/Wang2025, Sun2024, Baek2024} to address coverage limitations. However, pure generative approaches introduce new challenges: hallucination of non-existent or irrelevant queries, lack of grounding in actual searchable content, and generation of unsafe completions when not carefully controlled. Prior attempts at multi-objective QAC \cite{conf/emnlp/SinghFC24, Patki2024} apply diversity-aware losses or linear objective scalarization during retrieval, but still rely on ranking pre-generated candidates rather than leveraging the contextual reasoning capabilities of modern large language models (LLMs) for end-to-end list generation.

\textbf{Our Approach.} We propose a unified framework that reformulates QAC as direct list generation through Retrieval-Augmented Generation (RAG) and multi-objective alignment. Our key insight is that by generating complete suggestion lists in a single forward pass, conditioned on rich retrieved context, an LLM can simultaneously optimize for multiple objectives while ensuring groundedness and diversity and effectively generate suggestions for cases like those in Figure~\ref{fig:intro}. The framework consists of three main components: (1) a RAG-based pipeline that constructs structured prompts containing retrieved candidates, catalog metadata, and engagement signals; (2) a two-stage training procedure combining supervised fine-tuning on high-quality synthetic data and multi-objective Direct Preference Optimization (DPO) aligning the model with six objectives; (3) a hybrid serving architecture that balances quality and latency by using a Large Generator to pre-compute suggestions offline and a Compact Generator to handle uncached prefixes under strict latency constraints. 

We validate our framework through comprehensive evaluation on a large-scale commercial search platform. Offline experiments demonstrate substantial improvements over production baselines across relevance, safety, groundedness, and diversity metrics. Human evaluation confirms strong user preference (+0.40 to +0.69). Most importantly, a controlled online experiment with 10\% of production traffic shows significant real-world impact: 5.44\% reduction in user keystrokes and 3.46\% increase in suggestion adoption rate.

Our main contributions are:
\begin{itemize}
    \item We reformulate QAC as end-to-end list generation with multi-objective optimization, enabling holistic optimization across competing objectives.
    \item We define and deploy a suite of rule-based, model-based, and LLM-as-judge verifiers tailored to QAC, providing reward components for relevance, safety, groundedness, engagement, diversity, and format compliance.
    \item We present a comprehensive methodology that combines RAG, multi-objective DPO, and iterative critique-revision for synthetic data generation, using the verifier suite to guide alignment.
    \item We validate our approach through extensive evaluation on a large-scale production search system, demonstrating substantial improvements in offline metrics, human evaluation, and online A/B testing.
\end{itemize}

\section{Problem Formulation}
\label{sec:problem_formulation}

Query Auto-Completion aims to assist users in formulating search queries efficiently. Given a user-typed prefix $p$, the system should generate a list of suggestions that accurately predict the user's intent. We frame QAC as a direct, full-page generation problem rather than a retrieve-and-rank task.

\textbf{Formal Definition.} Let $\mathcal{P}$ denote the space of possible prefixes and $\mathcal{Q}$ the space of complete queries. Our objective is to learn a model $\mathcal{M}: \mathcal{P} \times \mathcal{C} \to \mathcal{Q}^k$ that generates an ordered list of $k$ suggestions $S = (q_1, q_2, \dots, q_k)$ maximizing user utility:
\begin{equation}
S^* = \underset{S \in \mathcal{Q}^k}{\operatorname{argmax}} \, U(S | p, C, E)
\end{equation}
where $C$ represents external context (retrieved candidates, catalog metadata, historical features detailed in Section~\ref{sec:prompt}) and $E$ denotes the search engine backend.

\textbf{Multi-Objective Utility.} Defining user utility precisely is inherently complex and domain-dependent. In our production setting, we decompose $U$ into six objectives that together characterize a helpful, safe, and grounded QAC system: \emph{relevance}, measuring how well suggestions cover plausible intents for the prefix; \emph{safety}, requiring avoidance of unsafe or policy-violating completions; \emph{engagement}, capturing the likelihood that a suggestion leads to downstream actions such as clicks or downloads; \emph{catalog groundedness}, ensuring each suggestion corresponds to actual searchable content and does not lead to empty result pages; \emph{context groundedness}, requiring suggestions to be supported by the retrieved context $C$ rather than the model's parametric memory; and \emph{diversity}, encouraging coverage of multiple intents while avoiding near-duplicate suggestions within the list. Our goal is to train a model that jointly optimizes these six objectives via multi-objective alignment, operationalized through a suite of verifiers and a composite reward signal used during training (Section~\ref{sec:dpo}).

\begin{figure*}[t]
  \centering
  \includegraphics[width=0.95\textwidth]{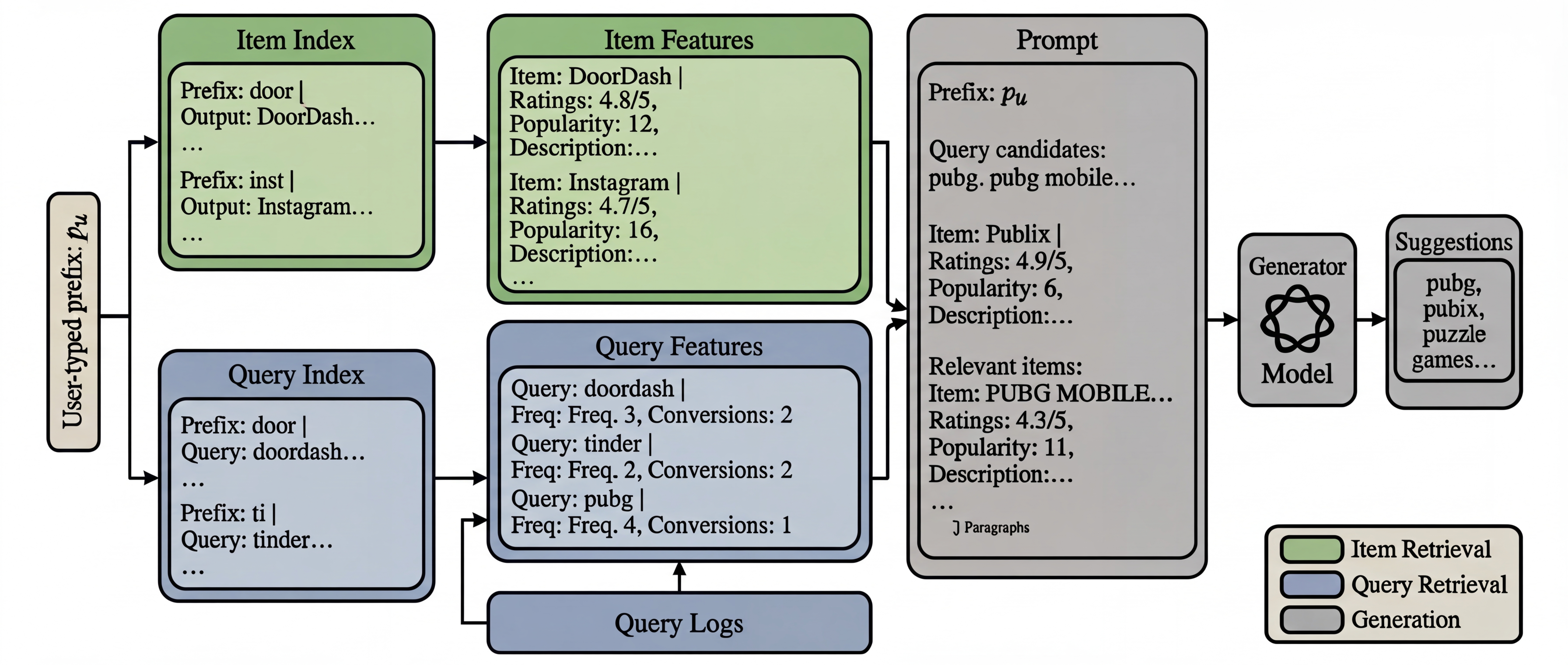}
  \caption{Overall RAG-based architecture for generating query auto-complete suggestions. Given a user prefix, the system retrieves information from multiple sources (query and search index, catalog, and query logs) to construct a prompt containing query candidates and relevant items with their features, which is then fed to the Generator to generate the final suggestions.}
  \label{fig:architecture}
\end{figure*}

\section{Methodology}
\label{methodology}
Our proposed QAC system is designed as a multi-stage, Retrieval-Augmented Generation (RAG) pipeline. The overall architecture is depicted in Figure~\ref{fig:architecture}. The system first generates a set of candidate queries from multiple sources, augments them with relevant contextual information, and then uses a fine-tuned generative model to produce the final list of suggestions. The generative model itself is tuned using a multi-objective framework, as shown in Figure~\ref{fig:tuning}, to align its outputs with several quality dimensions such as relevance, safety, and diversity.

\subsection{Architecture}
\label{sec:architecture}
The inference process begins when a user types a prefix. The system follows a structured path to generate a full list of QAC suggestions, as detailed below and illustrated in Figure~\ref{fig:architecture}.

\subsubsection{Candidate Generation}
To ensure both high recall and relevance, we generate an initial set of candidate queries from two complementary sources:

\begin{itemize}
    \item \textbf{Query Index:} A fast lookup table derived from historical search logs that maps common prefixes to frequently completed queries.

    \item \textbf{Content Retriever:} A search system combining textual and embedding-based retrieval with a learned ranker to handle novel, long-tail, or misspelled prefixes. We send the user's prefix to the Content Retriever and it returns items that would potentially satisfy the user's intent. This component complements the \texttt{Query Index} and prevents hallucinations by grounding suggestions in actual catalog items.
\end{itemize}

\subsubsection{Retrieval and Prompt Construction}
\label{sec:prompt}
We retrieve contextual information for all candidates to ground the final generation: \textbf{Catalog Metadata} (titles, descriptions, ratings, popularity) for each item, and \textbf{Query Features} (frequency, conversion rate) for each query. Exposing item metadata in the prompt lets the \texttt{Generator} generate suggestions grounded to (and linkable to) retrieved catalog items, reducing hallucinations. All collected information—user prefix, candidates, metadata, and features—is compiled into a single comprehensive \texttt{prompt}.

\subsubsection{Suggestion Generation}
The structured \texttt{prompt} is fed into a large language model referred to as the \texttt{Generator}. This model is tasked with generating the final, ordered list of QAC suggestions. Unlike traditional systems that simply rank a fixed set of candidates, the \texttt{Generator} synthesizes the information in the prompt to generate a complete, coherent, and optimized list of suggestions in a single pass. The fine-tuning process for this model, which enables it to balance multiple objectives, is detailed in the following section.

\subsection{Generator Model Fine-Tuning}
\label{sec:tuning}
\begin{figure*}[t]
  \centering
  \includegraphics[width=0.95\textwidth]{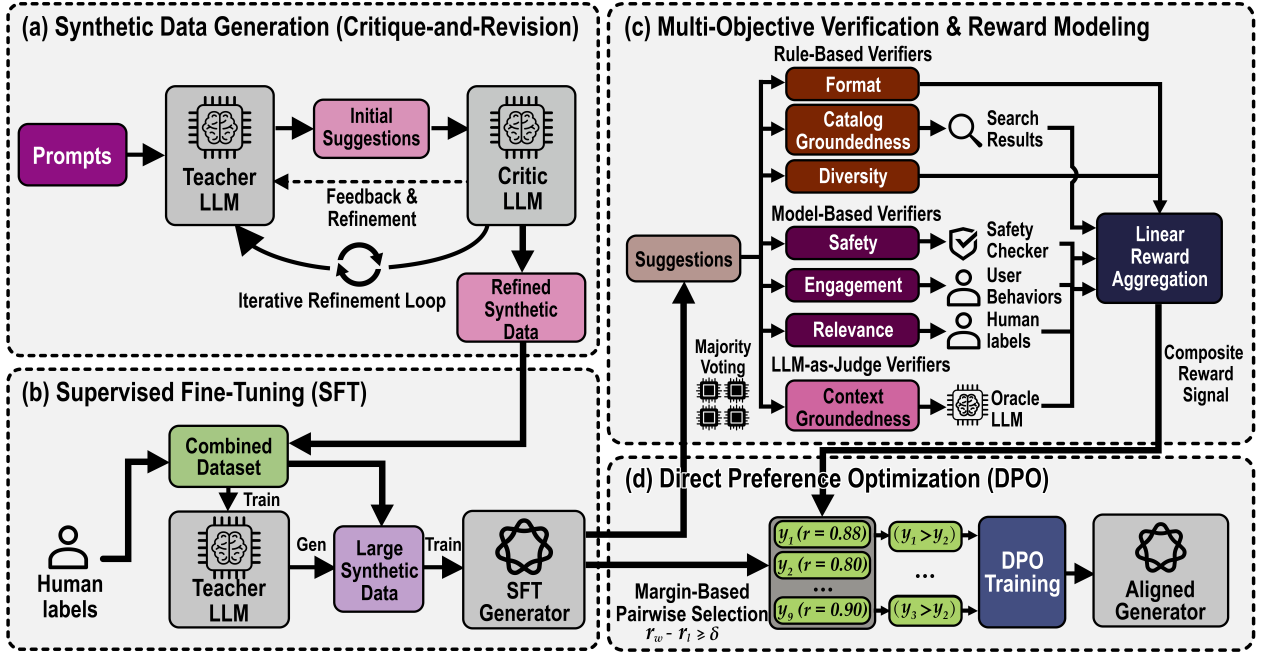}
  \caption{Training pipeline for the Generator model. (a) Critique-and-revision: a teacher LLM generates initial suggestion lists, which are iteratively refined based on feedback from a critic LLM to produce high-quality synthetic data. (b) Supervised Fine-Tuning (SFT): the Generator is trained on a mixture of human-labeled and synthetic examples. (c) Verifier training: rule-based, model-based, and LLM-as-judge verifiers are trained or defined for relevance, safety, catalog and context groundedness, engagement, diversity, and format. (d) Direct Preference Optimization (DPO): the SFT-trained Generator produces suggestion lists that are scored by the verifier suite; preference pairs constructed from the composite reward $R(p,S)$ are used to align the Generator with the multi-objective utility.}
  \label{fig:tuning}
\end{figure*}

To ensure the \texttt{Generator} produces high-quality suggestion lists, we employ a multi-stage training strategy, as illustrated in Figure~\ref{fig:tuning}: (1) high-quality synthetic data generation through an iterative critique-and-revision process with a teacher LLM; (2) supervised fine-tuning (SFT) of the Generator on synthetic and human-labeled data; (3) training a suite of verifiers that operationalize our multi-objective utility; and (4) multi-objective preference optimization via Direct Preference Optimization (DPO) using verifier-based rewards. All models are initialized from a proprietary foundation model pre-trained on web-scale text data. Appendix~\ref{app:gen-prompt}--\ref{app:revise-prompt} provide illustrative prompt templates for the Generator, Critic, and Reviser used in this pipeline.

\subsubsection{Supervised Fine-Tuning (SFT)}
\label{sec:sft}
The initial training phase uses Supervised Fine-Tuning to teach the \texttt{Generator} the task structure and desired output format. A key challenge in QAC is the scarcity of high-quality labeled data at scale. To address this, we develop a synthetic data generation pipeline based on a \texttt{teacher LLM}—a state-of-the-art large language model with strong instruction-following capabilities. Note that although we were able to leverage the teacher LLM for synthetic data generation, it was not suitable for deployment in our application due to the size and cost of the model. 

\textbf{Critique-and-Revision Process.} Simply prompting the teacher LLM to generate suggestions often produces suboptimal outputs that lack diversity or contain subtle quality issues. We introduce an iterative critique-and-revision procedure (illustrated in Figure~\ref{fig:tuning}):
\begin{enumerate}
    \item The teacher LLM generates initial suggestion lists for a diverse set of prefixes with varying popularity and complexity.
    \item A critic LLM analyzes each generated list and provides structured feedback on potential issues: semantic redundancy, poor relevance-diversity trade-offs, spelling inconsistencies, or lack of coverage across user intents.
    \item The teacher LLM revises its output based on the critique, producing refined suggestions.
\end{enumerate}


This process repeats for multiple rounds until the outputs meet quality criteria or subsequent revisions change only minimally. We then augment this synthetic data with a small set of human-labeled examples (approximately 10\% of the training set) and use the combined dataset to fine-tune the teacher LLM itself. The resulting tuned teacher model generates the final large-scale training corpus of approximately 50K prompt-suggestion pairs.

Formally, given this training dataset $\mathcal{D}_{\text{SFT}} = \{(x^{(i)}, y^{(i)})\}_{i=1}^{N_{\text{SFT}}}$ where $x^{(i)}$ represents the input prompt and $y^{(i)} = (y_1^{(i)}, \dots, y_{T_i}^{(i)})$ is the oracle-generated suggestion list tokenized into $T_i$ tokens, we minimize the standard language modeling loss:

\begin{equation}
\label{eq:sft-loss}
\mathcal{L}_{\text{SFT}}(\theta) = -\frac{1}{N_{\text{SFT}}} \sum_{i=1}^{N_{\text{SFT}}} \frac{1}{T_i} \sum_{t=1}^{T_i} \log P_\theta\left(y_t^{(i)} \mid y_{<t}^{(i)}, x^{(i)}\right)
\end{equation}

where $\theta$ denotes model parameters, $y_{<t}^{(i)}$ represents all preceding tokens, and $P_\theta$ is the predicted probability distribution. This trains the \texttt{Generator} to autoregressively produce well-structured suggestion lists conditioned on the prompt.

\subsubsection{Verifiers}
\label{sec:verifiers}
To operationalize the multi-objective utility in Section~\ref{sec:problem_formulation}, we define a suite of verifiers that score each generated suggestion list $S = (q_1,\dots,q_k)$ along the six objectives, plus an additional format constraint.

\textbf{Format verifier:} A rule-based format verifier checks that the Generator outputs a compact block that can be parsed directly by the serving system: an opening tag, one query per line, and a closing tag, for example
\texttt{<answer>} on the first line, followed by newline-separated queries, and \texttt{</answer>} on the last line. The verifier returns a binary indicator $I_{\text{fmt}}(S) \in \{0,1\}$, where $I_{\text{fmt}}(S) = 1$ if and only if the structure is valid and extraneous tokens are minimal. This indicator will later \emph{multiply} the composite reward so that misformatted outputs receive zero effective reward.
	
\textbf{Relevance verifier:} Let $R_{\text{rel}}(p, S)$ denote the relevance score for suggestion list $S$ given prefix $p$. A relevance verifier is fine-tuned using human-labeled judgments to score each individual suggestion $q_i$ against the prefix. Individual scores are aggregated using position-weighted discounting analogous to DCG \cite{Kalervo2002ndcg}.
	
\textbf{Engagement verifier:} Let $R_{\text{eng}}(p, S)$ quantify expected user engagement. In principle, this verifier can use any verifiable signal of user engagement, such as clicks, conversions, or other downstream actions. In our deployed setting, we instantiate it using a combination of (1) the conditional conversion probability given prefix $p$ and (2) the general historical conversion rate, aggregated across suggestions with positional discounting.
	
\textbf{Safety verifier:} Let $R_{\text{safe}}(S)$ measure the safety of suggestion list $S$. A safety verifier, trained on human-labeled data following our domain-specific safety policy, provides a binary judgment of whether each query $q_i$ is safe. A list is considered unsafe if any constituent query is flagged unsafe, and this signal is used both during data filtering and at alignment time.
	
\textbf{Catalog groundedness verifier:} We use $R_{\text{srg}}(S)$ as a rule-based catalog groundedness score, treating the production search engine backend as a catalog groundedness verifier: each query must return non-empty, reasonable search results. This encourages suggestions that correspond to real, searchable items and discourages hallucinations that lead to empty result pages.
	
\textbf{Context groundedness verifier:} We use $R_{\text{cg}}(p, S)$ as a context groundedness score. A context groundedness verifier built from LLM-as-judge with majority voting, trained on data derived from the teacher LLM and human-validated examples, predicts whether each query $q_i$ is derivable from the retrieved context $C$. This penalizes hallucinations that ignore or contradict the provided evidence.
	
\textbf{Diversity verifier:} Let $R_{\text{div}}(S)$ measure list-level semantic diversity by considering the distinctness of search result pages returned by each query $q_i$. We use an adjusted entropy formula that balances result distribution evenness with a penalty for redundancy:

\begin{align}
H_{\text{adj}} &=
\frac{
  H_{\text{st}}
  -
  \sum_{i:\, c_i > 1}
    \left[
      p_i^{(w)} \log_2\!\frac{1}{p_i^{(w)}}
      \cdot
      \frac{c_i}{T}
    \right]
}{
  \log_2 n
}
\label{eq:adjusted-entropy-final} \\[4pt]
H_{\text{st}} &=
-\sum_{i=1}^{n} p_i^{(w)} \log_2 p_i^{(w)} \\[4pt]
p_i^{(w)} &=
\frac{
  C_i / \log_2(i + 1)
}{
  \sum_{j=1}^{n} C_j / \log_2(j + 1)
}
\end{align}

where $H_{\text{st}}$ is standard entropy with position-weighted probabilities $p_i^{(w)}$, $H_{\text{adj}}$ is adjusted entropy penalizing results appearing in multiple suggestions, $C_i$ is the occurrence count of result $i$, $c_i$ is the number of suggestions containing result $i$, $T$ is the total number of suggestions, and $n$ is the total number of distinct results.



\subsubsection{Reward}
\label{sec:reward}
These individual scores are combined into a composite reward used for preference optimization. We first define a base linear reward $r_{\text{base}}(p, S)$ as a weighted sum of the continuous verifier outputs, and then gate it by the format indicator:
\begin{equation}
\label{eq:reward}
\begin{split}
r_{\text{base}}(p, S) &= w_{\text{rel}} \cdot R_{\text{rel}}(p, S) + w_{\text{eng}} \cdot R_{\text{eng}}(p, S) + w_{\text{safe}} \cdot R_{\text{safe}}(S) \\
&\quad + w_{\text{srg}} \cdot R_{\text{srg}}(S) + w_{\text{cg}} \cdot R_{\text{cg}}(p, S) + w_{\text{div}} \cdot R_{\text{div}}(S), \\
R(p,S) &= I_{\text{fmt}}(S) \cdot r_{\text{base}}(p, S),
\end{split}
\end{equation}
where the weight hyperparameters $\{w_j\}_{j=1}^{6}$ allow us to balance the trade-offs between the different objectives. These weights are tuned based on the specific business needs and user experience goals. Note that some rewards depend on the prefix $p$ (relevance, engagement, context groundedness), while others depend only on the suggestion list $S$ (safety, catalog groundedness, diversity). The multiplicative gate $I_{\text{fmt}}(S)$ ensures that misformatted outputs receive zero effective reward even if individual objective scores are high.
	
\subsubsection{Direct Preference Optimization (DPO)}
\label{sec:dpo}
Following SFT and verifier training, we further refine the model using Direct Preference Optimization (DPO) \cite{RafailovSMMEF23dpo}. This phase aligns the model's generated suggestion lists with the composite reward $R(p,S)$ by training it to prefer lists with higher reward scores over those with lower scores. To generate high-quality preference pairs, we employ a margin-based ranking strategy per prefix. For each prefix, we sample several candidate suggestion lists, score each using Equation~\ref{eq:reward}, and filter pairs to ensure meaningful quality distinctions: $r_{\text{chosen}} - r_{\text{rejected}} \ge \delta$ (typically $\delta = 0.08$–$0.10$). Valid pairs are ranked by margin in descending order, and we select the top-$k$ pairs (typically $k = 4$) per prefix. This approach captures diverse quality aspects while preventing over-representation and balancing dataset size with input coverage.
	
Given a preference dataset $\mathcal{D}_{\text{pref}} = \{(x^{(i)}, y_w^{(i)}, y_l^{(i)})\}_{i=1}^{M}$ where $x^{(i)}$ is the input prompt, $y_w^{(i)}$ is the chosen suggestion list (with higher reward), and $y_l^{(i)}$ is the rejected suggestion list (with lower reward), the DPO objective is:
	
\begin{equation}
\label{eq:dpo-loss}
\mathcal{L}_{\text{DPO}}(\theta) = -\mathbb{E}_{\mathcal{D}_{\text{pref}}} \left[ \log \sigma \left( \beta \log \frac{\pi_\theta(y_w|x)}{\pi_{\text{ref}}(y_w|x)} - \beta \log \frac{\pi_\theta(y_l|x)}{\pi_{\text{ref}}(y_l|x)} \right) \right]
\end{equation}
	
where $\pi_\theta$ is the \texttt{Generator} policy being optimized, $\pi_{\text{ref}}$ is the reference policy (the SFT model), $\beta$ controls the KL divergence constraint, and $\sigma$ is the sigmoid function. Here, $\pi_\theta(y|x) = \prod_{t=1}^{T} P_\theta(y_t | y_{<t}, x)$ denotes the autoregressive sequence probability under policy $\pi_\theta$. This objective increases the likelihood of high-reward suggestions while maintaining proximity to the reference model to prevent reward over-optimization.

\section{Offline Evaluation}
\label{sec:offline_eval}

We first evaluate our approach on a static dataset to measure quality across multiple dimensions without the complexity of production deployment.

\subsection{Experimental Setup}

\paragraph{Dataset}
There is no public large-scale dataset that matches our production QAC setting, so we evaluate on a proprietary dataset from a large-scale commercial search system. The offline evaluation set consists of 50{,}000 query prefixes sampled from real, anonymized user traffic. Prefixes are stratified to reflect the production distribution over popular (``head''), medium-frequency (``torso''), and rare (``tail'') prefixes.

\paragraph{Benchmark Methods}
We compare our full model against a production baseline and ablations:
\begin{itemize}
    \item \textbf{LTR Baseline:} A traditional two-stage system generating candidates from query logs and ranking them with a learning-to-rank model trained on lexical and engagement features.
    \item \textbf{LTR + Seq2Seq:} A hybrid approach combining the LTR baseline with a generative model trained on historical user queries to predict query completions given a prefix. When the LTR system produces fewer than the desired number of suggestions, the generative model produces additional individual queries to fill the list, providing improved coverage for long-tail prefixes.
    \item \textbf{SFT-only:} Generator fine-tuned only with SFT on oracle data (Section~\ref{sec:sft}), without DPO, to measure the impact of preference optimization.
    \item \textbf{SFT + DPO w/o Eng:} Generator fine-tuned with SFT and DPO but excluding the engagement reward, isolating the impact of historical user signals.
    \item \textbf{Full Model:} Complete model with SFT and DPO using all verifier-based reward components.
\end{itemize}
We instantiate two Generator variants: a \emph{Large Generator} used for high-quality offline pre-generation, and a \emph{Compact Generator} used for latency-critical online inference. Both share the same training pipeline; Section~\ref{sec:serving} describes how they are used in serving.

\paragraph{Metrics}
We measure performance against the individual verifier objectives (Section \ref{sec:verifiers}). For each prefix $p$ with traffic weight $w_p$ and generated suggestion list $S_p$, we compute traffic-weighted averages of per-prefix scores $m_p$:
\begin{equation}
\mathrm{Metric} = \frac{\sum_{p} w_p m_p}{\sum_{p} w_p},
\end{equation}
so that our offline metrics reflect the production request distribution.

\noindent\textbf{Coverage.}
Here $m_p = I_{\text{offer}}(p) \in \{0,1\}$ indicates whether the system returns at least one suggestion for prefix $p$, and higher values are preferred.

\noindent\textbf{Relevance.}
Here $m_p = R_{\text{rel}}(p, S_p) \in [0,1]$ is the page-level relevance score from the relevance verifier (Section~\ref{sec:verifiers}), averaged over prefixes with traffic weighting; higher is better.

\noindent\textbf{Unsafe rate.}
Here $m_p = I_{\text{unsafe}}(p) \in \{0,1\}$ indicates whether the safety verifier flags at least one suggestion in $S_p$ as unsafe; lower values are better.

\noindent\textbf{Engagement win rate.}
Here $m_p = I_{\text{eng-win}}(p) \in \{0,1\}$ indicates whether the engagement score from the engagement verifier (e.g., conditional conversion probability) exceeds the corresponding score for the LTR baseline; higher values are better.

\noindent\textbf{Catalog ungrounded rate.}
Here $m_p = I_{\text{cat-ung}}(p) \in \{0,1\}$ indicates whether at least one suggestion in $S_p$ is marked ungrounded by the catalog groundedness verifier (i.e., returns no acceptable results from the search backend); lower values are better.

\noindent\textbf{Context ungrounded rate.}
Here $m_p = I_{\text{ctx-ung}}(p) \in \{0,1\}$ indicates whether at least one suggestion in $S_p$ is marked ungrounded by the context groundedness verifier (LLM-as-judge); lower values are better.

\noindent\textbf{Diversity.}
Here $m_p = R_{\text{div}}(S_p)$ is the page-level diversity score defined in Equation~\ref{eq:adjusted-entropy-final}, with higher values indicating more diverse search result pages across suggestions.

\subsection{Results}

\begin{table*}[!t]
\centering
\caption{Offline evaluation results across different models. Our full model shows superior performance on all metrics.}
\label{tab:offline_results}
\begin{tabular}{lccccccc}
\toprule
\textbf{Model} & \textbf{Coverage} & \textbf{Relevance} & \textbf{UnsafeRate} & \textbf{EngWinRate} & \textbf{CatalogUngrdRate} & \textbf{CtxUngrdRate} & \textbf{Diversity} \\
\midrule
LTR Baseline & 84.6\% & 0.646 & \textbf{0.37\%} & - & \textbf{0.28\%} & - & 72.63 \\
LTR + Seq2Seq & \textbf{97.72\%} & 0.685 & 0.79\% & 14.47\% & 0.67\% & - & \textbf{76.99} \\
\hline
\multicolumn{8}{c}{\textit{Large Generator}} \\
\hline
SFT-only & 93.8\% & 0.685 & 1.01\% & 2.37\% & 3.96\% & 9.72\% & 75.78 \\
SFT+DPO w/o Eng & 93.1\% & \textbf{0.690} & 0.64\% & -1.32\% & 0.56\% & 10.54\% & 74.06 \\
Full & 93.0\% & 0.687 & 0.65\% & \textbf{6.58\%} & 0.49\% & \textbf{9.43\%} & 74.10 \\
\hline
\multicolumn{8}{c}{\textit{Compact Generator}} \\
\hline
SFT-only & 94.1\% & 0.677 & 1.04\% & 2.37\% & 0.85\% & 16.42\% & 73.95 \\
SFT+DPO w/o Eng & 93.5\% & \textbf{0.687} & 0.77\% & -6.16\% & 0.64\% & \textbf{14.90\%} & 73.60 \\
Full & 93.7\% & 0.682 & 0.72\% & \textbf{16.28\%} & 0.59\% & 15.85\% & 74.16 \\
\bottomrule
\end{tabular}
\end{table*}

Table~\ref{tab:offline_results} presents offline evaluation results across all models. Our methods demonstrate substantially improved \textit{Relevance} compared to the LTR Baseline, with all DPO-tuned variants achieving scores above 0.68 versus 0.646 for the baseline.

A key trade-off emerges between coverage and content quality. Generative models that can predict queries for arbitrary prefixes achieve substantially higher \textit{Coverage} than the LTR Baseline (97.72\% for LTR + Seq2Seq, 93.0–94.1\% for our models, versus 84.6\% for LTR Baseline), but at the cost of higher \textit{UnsafeRate} and \textit{CatalogUngrdRate}. Our DPO-tuned models achieve competitive Coverage (93.0–93.7\%) while substantially improving safety compared to LTR + Seq2Seq: the Full model in the Large Generator block achieves 0.65\% UnsafeRate versus 0.79\%, and 0.49\% CatalogUngrdRate versus 0.67\%. This demonstrates that multi-objective DPO alignment can maintain high coverage while effectively mitigating the safety risks inherent in generative approaches.

Our models also achieve the best \textit{Diversity} scores among trainable systems (74–76 range), approaching the LTR + Seq2Seq baseline's 76.99 while maintaining superior relevance and safety. The \textit{CtxUngrdRate} results validate our RAG-based approach, with the Full model in the Large Generator block achieving a 9.43\% context ungrounded rate—the best among models with available metrics.

\subsection{Ablation Study}

To understand the contribution of each component in our training pipeline, we conduct ablation studies comparing models trained with different configurations.

\paragraph{Impact of DPO Fine-Tuning}
Comparing \textbf{SFT-only} to \textbf{SFT + DPO w/o Eng} reveals the substantial impact of multi-objective preference tuning. While SFT-only produces syntactically correct lists with high \textit{Coverage} (93.8–94.1\%), it fails to balance the competing objectives of Relevance, Diversity, and safety. The DPO tuning step maintains comparable \textit{Coverage} (93.1–93.5\%) while substantially improving safety and catalog groundedness: in the Large Generator, DPO reduces \textit{UnsafeRate} from 1.01\% to 0.64\% and \textit{CatalogUngrdRate} from 3.96\% to 0.56\%. Similar improvements are observed in the Compact Generator. This systematic improvement demonstrates that DPO alignment is critical for producing high-quality, safe, and well-grounded suggestions at scale, with only a minimal trade-off in Coverage.

\paragraph{Impact of Engagement Reward}
Comparing \textbf{SFT + DPO w/o Eng} to \textbf{Full Model} isolates the contribution of historical user interaction signals. Adding engagement as an objective yields dramatic improvements in \textit{EngWinRate}—from -1.32\% to +6.58\% for the Large Generator and from -6.16\% to +16.28\% for the Compact Generator—while maintaining near on-par performance on \textit{Relevance} (0.690 vs 0.687 for the Large Generator, 0.687 vs 0.682 for the Compact Generator). This validates that incorporating historical engagement data enables the model to better prioritize suggestions that lead to successful user journeys, reflected in substantially higher suggestion click-through rates, without sacrificing the core Relevance objective.

\section{Serving}
\label{sec:serving}

\begin{figure}[t]
  \centering
  \vspace*{-0.5cm}
  \includegraphics[width=0.48\textwidth]{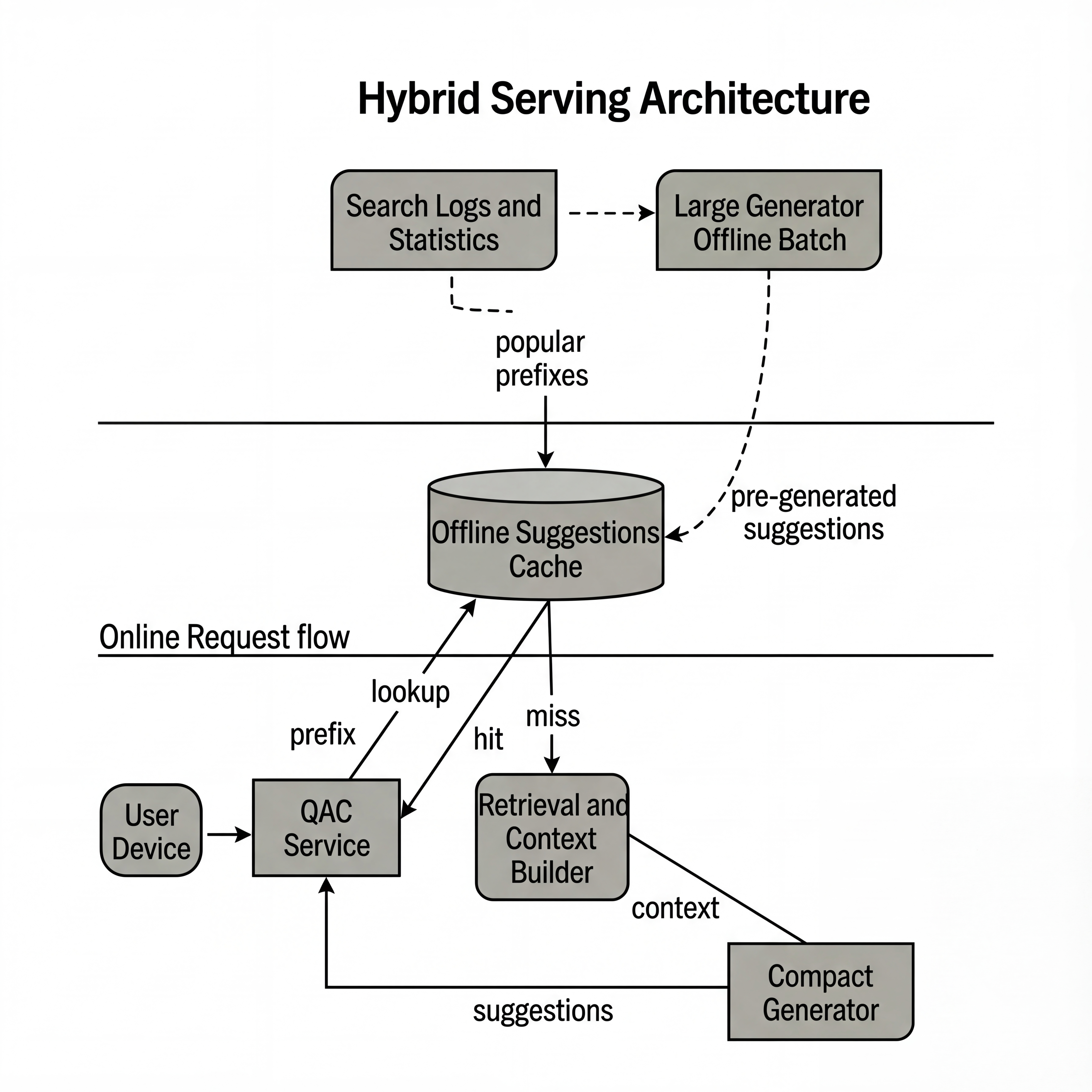}
  \caption{Hybrid serving architecture. Frequently requested prefixes are pre-generated offline by the Large Generator and stored in a prefix cache that the QAC service queries first, serving most requests without online model latency. Cache misses trigger real-time retrieval and generation with the Compact Generator via the Retrieval and Context Builder, ensuring coverage for long-tail and emerging prefixes.}
  \label{fig:serving}
\end{figure}

Query auto-completion requires fast response times to provide seamless user experience. While our Large Generator achieves superior quality, it has higher inference latency. The Compact Generator reduces latency with slightly lower standalone quality. We employ a hybrid two-tier serving architecture that balances quality and latency, illustrated in Figure~\ref{fig:serving}.

\textbf{Offline Batch Inference.} For high-frequency prefixes, we pre-compute suggestions using the Large Generator with no latency constraints. Analysis of search logs identifies the most common prefixes. These suggestions are generated offline and cached for online serving, enabling large-model results for the majority of requests with only sub-millisecond online latency dominated by key-value lookup.

\textbf{Online Real-time Inference.} For cache misses—typically long-tail, rare, or newly trending prefixes—we perform real-time inference using the Compact Generator. Despite its smaller size, the Compact Generator maintains strong quality, achieving acceptable performance for low-frequency queries with end-to-end latency on the order of one hundred milliseconds, depending on retrieval cost, inference hardware, and prefix-level caching.

\section{Human Evaluation}
\label{sec:human_eval}

To complement automated metrics, we conduct human evaluations to measure perceived quality and user preference for our generated suggestions.

\subsection{Evaluation Setup}
We run two complementary studies. (1) \textbf{Item-wise Relevance:} raters score individual suggestions on a predefined scale, which we aggregate with positional and traffic weighting into a page-level score in $[0,1]$ (higher is better). (2) \textbf{Pairwise Preference:} raters compare two suggestion lists side-by-side and indicate preference strength (neutral $=0$, slight $=+1$, strong $=+2$), capturing holistic aspects such as diversity and usefulness.

Both evaluations were conducted on our production deployment system using the hybrid serving architecture in Section~\ref{sec:serving}. For each method (e.g., SFT-only, SFT + DPO w/o Eng, Full), the Large Generator produced offline batches of candidate suggestions, while the Compact Generator handled real-time inference for cache misses. This mirrors the user-facing configuration, so raters evaluated the same behavior users see in production.

\subsection{Results}
Table~\ref{tab:human_eval} summarizes the results. The Full model attains the highest item-wise relevance score of 0.699, substantially outperforming the LTR Baseline's 0.653. Scores progressively improve from SFT-only (0.689) to SFT + DPO w/o Eng (0.698) to Full (0.699), indicating that each training stage contributes to perceived quality.

Pairwise preference shows a complementary view: SFT + DPO w/o Eng achieves the highest preference score ($+0.69$), while the Full model is slightly lower ($+0.40$). This suggests that adding engagement as an objective introduces mild qualitative trade-offs that raters notice in side-by-side comparisons, even as it improves user interaction metrics. All aligned models nonetheless outperform the LTR Baseline.

\begin{table}[t]
\centering
\caption{Human evaluation results using hybrid deployment (Large Generator offline + Compact Generator online). Our full model achieves the highest item-wise relevance score.}
\label{tab:human_eval}
\begin{tabular}{lcc}
\toprule
\textbf{Method} & \textbf{Item-wise Score} & \textbf{Pairwise Pref.} \\
\midrule
LTR Baseline & 0.653 & - \\
\hline
SFT-only & 0.689 & $+0.5$ \\
SFT + DPO w/o Eng & 0.698 & $\mathbf{+0.69}$ \\
\textbf{Full} & \textbf{0.699} & $+0.40$ \\
\bottomrule
\end{tabular}
\end{table}

\section{Online A/B Test}
\label{sec:online_test}

We deployed our full model with hybrid inference in a live A/B test against the existing \textbf{Production Baseline}, routing 10\% of production traffic to the treatment. As shown in Table~\ref{tab:ab_results}, our approach achieves a 5.44\% reduction in characters typed and a 3.46\% increase in suggestions taken (CTR), indicating that users formulate queries with less effort and place greater trust in the provided completions.

\begin{table}[ht]
\centering
\caption{Online A/B test results comparing our full model against the production baseline. All improvements are statistically significant (p < 0.05).}
\label{tab:ab_results}
\begin{tabular}{lcc}
\toprule
\textbf{Metric} & \textbf{Relative Change} \\
\midrule
Characters Typed & -5.44\% \\
Suggestions Taken (CTR) & +3.46\% \\
\bottomrule
\end{tabular}
\end{table}

These results demonstrate that our framework not only improves offline and human-rated quality but also translates to a more efficient and engaging user experience in production.

\section{Related Work}
\label{rel_work}

\textbf{Retrieve-and-Rank QAC Systems.} Traditional QAC follows a two-stage pipeline \cite{Cai2016ASO, kai2021, Hasan2011}: candidate generation from query logs followed by learning-to-rank \cite{BarYossef2011, conf/sigir/Zhang2015, conf/cikm/Wang2020}. While effective for common queries, these systems struggle with long-tail prefixes absent from historical logs \cite{fiorini2018-personalized}, require extensive feature engineering, and cannot generate novel suggestions \cite{verma2023, Gog2020}.

\textbf{Generative QAC.} To address coverage limitations, researchers have explored generative models including RNNs \cite{Park2017, Sordoni2015, Jiang2018}, LSTMs \cite{jaech-2018-personalized, Cheng2021}, and transformers \cite{Mustar2020, Mustar2021}. Recent work applies large language models \cite{conf/coling/LiZXLW25, conf/naacl/DholeBCA24, conf/kdd/Wang2025}, which can generalize to unseen prefixes. However, these approaches still face hallucination, lack of grounding in searchable content, and difficulty optimizing multiple objectives simultaneously. Our work combines generation with retrieval augmentation and explicit multi-objective alignment to address these challenges.

\textbf{Multi-Objective QAC.} Several efforts have tackled multiple objectives in QAC, primarily focusing on diversity alongside relevance. The DiAL framework \cite{conf/emnlp/SinghFC24} optimizes a smooth approximation of $\alpha$NDCG as a listwise ranking loss, while MONR \cite{Patki2024} proposes multi-objective neural retrieval. Wang et al. \cite{conf/kdd/Wang2025} apply RL to detoxify generated queries. Most similar to our work, Bodigutla et al. \cite{Bodigutla2021} combine naturalness, relatedness, and user feedback rewards for related query suggestions using RL. However, these approaches either operate at the retrieval stage (losing the benefits of LLM reasoning) or optimize only 2-3 objectives. We extend this line of work by jointly optimizing six objectives—relevance, safety, engagement, catalog groundedness, context groundedness, and diversity—in an end-to-end generation framework.

\textbf{RAG for QAC.} Retrieval-Augmented Generation \cite{LewisRag2020} has shown strong results across NLP tasks. For QAC specifically, Baek et al. \cite{Baek2024} use RAG for personalized next-query suggestions conditioned on web pages, while Sun et al. \cite{Sun2024} contextualize generation with retrieved product metadata. Our approach advances RAG-based QAC by: (1) retrieving richer context including candidates, catalog metadata, and engagement signals; (2) generating complete optimized lists rather than individual queries; (3) explicitly training for groundedness through dedicated reward components.

\textbf{Alignment Methods.} Recent advances in LLM alignment have moved from Reinforcement Learning from Human Feedback (RLHF) \cite{Christiano2017rlhf} and Proximal Policy Optimization (PPO) \cite{Schulman2017ppo, Ziegler2019, Stiennon2020} toward more efficient methods like Direct Preference Optimization (DPO) \cite{RafailovSMMEF23dpo} and its variants \cite{Ethayarajh2024}. While these methods have been applied to general text generation tasks, their application to multi-objective QAC with explicit groundedness constraints remains unexplored. We demonstrate that DPO can effectively align a generative QAC model with six diverse objectives, combining model-based judges (relevance, safety, context groundedness) with rule-based metrics (engagement, catalog groundedness, diversity).

\textbf{End-to-End Retrieval--Generation Optimization.} A complementary line of work explores jointly optimizing retrievers and generators with reinforcement learning, using verifiers or task rewards as signals \cite{aisearch2025}. Our current framework keeps the retrieval stack fixed and focuses alignment on the Generator via multi-objective DPO using verifier-derived rewards. Extending this to an end-to-end setting---where both retrieval and generation are optimized jointly with verifier-only rewards---is a promising direction for future QAC systems.

\textbf{Our Contributions.} We advance the state-of-the-art in query auto-completion through three key contributions. First, we reformulate QAC as end-to-end list generation with multi-objective optimization, enabling more holistic optimization than traditional multi-stage retrieve-and-rank pipelines. Second, we present a comprehensive methodology combining RAG, multi-objective DPO with carefully designed learned and rule-based verifiers, and iterative critique-revision for synthetic data quality. Third, we provide rigorous validation through offline metrics, human evaluation, and a large-scale online experiment in a production search system, demonstrating substantial improvements while maintaining safety and groundedness guarantees.

\section{Conclusion}
\label{conclusion}

We presented a unified framework for Query Auto-Completion that reformulates the task as end-to-end list generation with multi-objective optimization. Our methodology combines Retrieval-Augmented Generation, multi-objective Direct Preference Optimization with rule-based, model-based, and LLM-as-judge verifiers, and iterative critique-revision for synthetic data generation. Comprehensive evaluation demonstrates substantial improvements: offline metrics show consistent gains across all six objectives, human evaluation yields +0.40 to +0.69 preference scores over baselines, and a controlled online experiment with 10\% of production traffic achieves a 5.44\% reduction in characters typed and a 3.46\% increase in suggestion adoption rate.

Our results establish that end-to-end generation with multi-objective alignment provides a principled and effective approach to production QAC systems. The framework successfully addresses the fundamental trade-offs in QAC—relevance, safety, groundedness, engagement, and diversity—while meeting strict latency requirements via a hybrid serving architecture that combines offline Large Generator pre-computation with a Compact Generator for real-time inference. Because the approach only assumes a search backend, a query index, and modest human labels for verifiers, it is applicable beyond mobile app search to other domains such as e-commerce, media, and content recommendation where users rely on query suggestions to explore large catalogs.

Future work includes exploring personalization through user- and session-specific context encoding, extending the framework to multi-modal search domains, and moving toward joint optimization of retrieval and generation with verifier-based rewards, as well as investigating alternative alignment objectives and verifier designs tailored to domain-specific requirements.

\begin{acks}
We thank Roberto Konow for support and contributions to this research and its productionization. We also thank Evangelia Christakopoulou for peer review and helpful comments, and Matt Jocker for peer review and publication coordination.
\end{acks}

\bibliographystyle{ACM-Reference-Format}
\bibliography{references}

@inproceedings{conf/emnlp/SinghFC24,
  author = {Singh, Sonali and Farfade, Sachin and Comar, Prakash Mandayam},
  booktitle = {EMNLP (Industry Track)},
  ee = {https://aclanthology.org/2024.emnlp-industry.87},
  isbn = {979-8-89176-166-7},
  pages = {1152-1162},
  publisher = {Association for Computational Linguistics},
  title = {DiAL : Diversity Aware Listwise Ranking for Query Auto-Complete.},
  year = 2024
}

@inproceedings{Patki2024,
 author = {Rohit Patki and Sravan Bodapati and Christopher Potts},
 title = {Multi-objective neural retrieval for query autocomplete},
 booktitle = {WWW '23 Companion},
 year = {2023},
 url = {https://www.amazon.science/publications/multi-objective-neural-retrieval-for-query-autocomplete},
}

@inproceedings{conf/naacl/DholeBCA24,
  author = {Dhole, Kaustubh D. and Bajaj, Shivam and Chandradevan, Ramraj and Agichtein, Eugene},
  booktitle = {NAACL (Demonstrations)},
  isbn = {979-8-89176-116-2},
  pages = {107-115},
  publisher = {Association for Computational Linguistics},
  title = {QueryExplorer: An Interactive Query Generation Assistant for Search and Exploration.},
  year = 2024
}

@article{conf/kdd/Wang2025,
  author = {Wang, Zhibo and Jiang, Xiaoze and Qin, Zhiheng and Yu, Enyun and Li, Han},
  journal = {KDD},
  title = {Personalized Query Auto-Completion for Long and Short-Term Interests with Adaptive Detoxification Generation.},
  year = 2025
}

@inproceedings{conf/coling/LiZXLW25,
  author = {Li, Zhipeng and Zheng, Shuang and Xiao, Jiaping and Li, Xianneng and Wang, Lei},
  booktitle = {COLING (Industry)},
  isbn = {979-8-89176-197-1},
  pages = {679-688},
  publisher = {Association for Computational Linguistics},
  title = {UCTG: A Unified Controllable Text Generation Framework for Query Auto-Completion.},
  year = 2025
}

@inproceedings{Baek2024,
    author = {Baek, Jinheon and Chandrasekaran, Nirupama and Cucerzan, Silviu and Herring, Allen and Jauhar, Sujay Kumar},
    title = {Knowledge-Augmented Large Language Models for Personalized Contextual Query Suggestion},
    year = {2024},
    isbn = {9798400701719},
    publisher = {Association for Computing Machinery},
    address = {New York, NY, USA},
    url = {https://doi.org/10.1145/3589334.3645404},
    doi = {10.1145/3589334.3645404},
    booktitle = {Proceedings of the ACM Web Conference 2024},
    pages = {3355–3366},
    location = {Singapore, Singapore},
    series = {WWW '24}
}

@inproceedings{conf/cikm/Wang2020,
    author = {Wang, Sida and Guo, Weiwei and Gao, Huiji and Long, Bo},
    title = {Efficient Neural Query Auto Completion},
    year = {2020},
    isbn = {9781450368599},
    publisher = {Association for Computing Machinery},
    address = {New York, NY, USA},
    url = {https://doi.org/10.1145/3340531.3412701},
    doi = {10.1145/3340531.3412701},
    booktitle = {Proceedings of the 29th ACM International Conference on Information \& Knowledge Management},
    pages = {2797–2804},
    location = {Virtual Event, Ireland},
    series = {CIKM '20}
}

@inproceedings{Park2017,
    author = {Park, Dae Hoon and Chiba, Rikio},
    title = {A Neural Language Model for Query Auto-Completion},
    year = {2017},
    isbn = {9781450350228},
    publisher = {Association for Computing Machinery},
    address = {New York, NY, USA},
    url = {https://doi.org/10.1145/3077136.3080758},
    doi = {10.1145/3077136.3080758},
    booktitle = {Proceedings of the 40th International ACM SIGIR Conference on Research and Development in Information Retrieval},
    pages = {1189–1192},
    numpages = {4},
    location = {Shinjuku, Tokyo, Japan},
    series = {SIGIR '17}
}

@inproceedings{conf/sigir/Zhang2015,
    author = {Zhang, Aston and Goyal, Amit and Kong, Weize and Deng, Hongbo and Dong, Anlei and Chang, Yi and Gunter, Carl A. and Han, Jiawei},
    title = {adaQAC: Adaptive Query Auto-Completion via Implicit Negative Feedback},
    year = {2015},
    isbn = {9781450336215},
    publisher = {Association for Computing Machinery},
    address = {New York, NY, USA},
    url = {https://doi.org/10.1145/2766462.2767697},
    doi = {10.1145/2766462.2767697},
    booktitle = {Proceedings of the 38th International ACM SIGIR Conference on Research and Development in Information Retrieval},
    pages = {143–152},
    location = {Santiago, Chile},
    series = {SIGIR '15}
}

@inproceedings{Sun2024,
    author = {Andy Sun and Tianqi Zheng and Aakash Kolekar and Rohit Patki and Hossein Khazaei and Xuan Guo and George Cai and David Liu and Ruirui Li and Yupin Huang and Dante Everaert and Hanqing Lu and Garima Patel and Monica Cheng},
    title = {A product-aware query auto-completion framework for e-commerce search via retrieval-augmented generation method},
    year = {2024},
    booktitle = {Proceedings of the Workshop Information Retrieval's Role in RAG Systems (IR-RAG 2024)}
}

@inproceedings{Santu2017,
    author = {Karmaker Santu, Shubhra Kanti and Sondhi, Parikshit and Zhai, ChengXiang},
    title = {On Application of Learning to Rank for E-Commerce Search},
    year = {2017},
    isbn = {9781450350228},
    publisher = {Association for Computing Machinery},
    address = {New York, NY, USA},
    url = {https://doi.org/10.1145/3077136.3080838},
    doi = {10.1145/3077136.3080838},
    booktitle = {Proceedings of the 40th International ACM SIGIR Conference on Research and Development in Information Retrieval},
    pages = {475–484},
    location = {Shinjuku, Tokyo, Japan},
    series = {SIGIR '17}
}

@inproceedings{BarYossef2011,
    title	= {Context-sensitive query auto-completion},
    author	= {Ziv Bar-Yossef and Naama Kraus},year	= {2011},
    booktitle	= {Proceedings of the 20th International Conference on World Wide Web (WWW)},
    pages	= {107--116}
}

@article{kai2021,
  author       = {Kai Yuan and Da Kuang},
  title        = {Deep Pairwise Learning To Rank For Search Autocomplete},
  journal      = {CoRR},
  volume       = {abs/2108.04976},
  year         = {2021},
  url          = {https://arxiv.org/abs/2108.04976},
  eprinttype    = {arXiv}
}

@inproceedings{Hasan2011,
    author = {Hasan, Mohammad Al and Parikh, Nish and Singh, Gyanit and Sundaresan, Neel},
    title = {Query suggestion for E-commerce sites},
    year = {2011},
    isbn = {9781450304931},
    publisher = {Association for Computing Machinery},
    address = {New York, NY, USA},
    url = {https://doi.org/10.1145/1935826.1935927},
    doi = {10.1145/1935826.1935927},
    booktitle = {Proceedings of the Fourth ACM International Conference on Web Search and Data Mining},
    pages = {765–774},
    location = {Hong Kong, China},
    series = {WSDM '11}
}

@inproceedings{Gog2020,
    author = {Gog, Simon and Pibiri, Giulio Ermanno and Venturini, Rossano},
    title = {Efficient and Effective Query Auto-Completion},
    year = {2020},
    isbn = {9781450380164},
    publisher = {Association for Computing Machinery},
    address = {New York, NY, USA},
    url = {https://doi.org/10.1145/3397271.3401432},
    doi = {10.1145/3397271.3401432},
    booktitle = {Proceedings of the 43rd International ACM SIGIR Conference on Research and Development in Information Retrieval},
    pages = {2271–2280},
    location = {Virtual Event, China},
    series = {SIGIR '20}
}

@inproceedings{fiorini2018-personalized,
    title = "Personalized neural language models for real-world query auto completion",
    author = "Fiorini, Nicolas  and Lu, Zhiyong",
    booktitle = "Proceedings of the 2018 Conference of the North {A}merican Chapter of the Association for Computational Linguistics: Human Language Technologies, Volume 3 (Industry Papers)",
    year = "2018",
    address = "New Orleans - Louisiana",
    publisher = "Association for Computational Linguistics",
    url = "https://aclanthology.org/N18-3026/",
    doi = "10.18653/v1/N18-3026",
    pages = "208--215",
}

@inproceedings{jaech-2018-personalized,
    title = "Personalized Language Model for Query Auto-Completion",
    author = "Jaech, Aaron  and Ostendorf, Mari",
    booktitle = "Proceedings of the 56th Annual Meeting of the Association for Computational Linguistics (Volume 2: Short Papers)",
    year = "2018",
    address = "Melbourne, Australia",
    publisher = "Association for Computational Linguistics",
    url = "https://aclanthology.org/P18-2111/",
    doi = "10.18653/v1/P18-2111",
    pages = "700--705"
}

@inproceedings{Sordoni2015,
    author = {Sordoni, Alessandro and Bengio, Yoshua and Vahabi, Hossein and Lioma, Christina and Grue Simonsen, Jakob and Nie, Jian-Yun},
    title = {A Hierarchical Recurrent Encoder-Decoder for Generative Context-Aware Query Suggestion},
    year = {2015},
    isbn = {9781450337946},
    publisher = {Association for Computing Machinery},
    address = {New York, NY, USA},
    url = {https://doi.org/10.1145/2806416.2806493},
    doi = {10.1145/2806416.2806493},
    booktitle = {Proceedings of the 24th ACM International on Conference on Information and Knowledge Management},
    pages = {553–562},
    location = {Melbourne, Australia},
    series = {CIKM '15}
}

@article{Cai2016ASO,
  title={A Survey of Query Auto Completion in Information Retrieval},
  author={Fei Cai and M. de Rijke},
  journal={Found. Trends Inf. Retr.},
  year={2016},
  volume={10},
  pages={273-363},
  url={https://api.semanticscholar.org/CorpusID:207178818}
}

@inproceedings{Cheng2021,
    author = {Cheng, Qiannan and Ren, Zhaochun and Lin, Yujie and Ren, Pengjie and Chen, Zhumin and Liu, Xiangyuan and de Rijke, Maarten de},
    title = {Long Short-Term Session Search: Joint Personalized Reranking and Next Query Prediction},
    year = {2021},
    isbn = {9781450383127},
    publisher = {Association for Computing Machinery},
    address = {New York, NY, USA},
    url = {https://doi.org/10.1145/3442381.3449941},
    doi = {10.1145/3442381.3449941},
    booktitle = {Proceedings of the Web Conference 2021},
    pages = {239–248},
    numpages = {10},
    location = {Ljubljana, Slovenia},
    series = {WWW '21}
}

@article{Mustar2021,
    author = {Mustar, Agn\`{e}s and Lamprier, Sylvain and Piwowarski, Benjamin},
    title = {On the Study of Transformers for Query Suggestion},
    year = {2021},
    issue_date = {January 2022},
    publisher = {Association for Computing Machinery},
    address = {New York, NY, USA},
    volume = {40},
    number = {1},
    issn = {1046-8188},
    url = {https://doi.org/10.1145/3470562},
    doi = {10.1145/3470562},
    journal = {ACM Trans. Inf. Syst.},
    month = oct,
    articleno = {18},
    numpages = {27}
}

@inproceedings{Mustar2020,
  author       = {Agn{\`{e}}s Mustar and
                  Sylvain Lamprier and
                  Benjamin Piwowarski},
  editor       = {Iv{\'{a}}n Cantador and
                  Max Chevalier and
                  Massimo Melucci and
                  Josiane Mothe},
  title        = {Using {BERT} and {BART} for Query Suggestion},
  booktitle    = {Proceedings of the First Joint Conference of the Information Retrieval
                  Communities in Europe {(CIRCLE} 2020), Samatan, Gers, France, July
                  6-9, 2020},
  series       = {{CEUR} Workshop Proceedings},
  volume       = {2621},
  publisher    = {CEUR-WS.org},
  year         = {2020},
  url          = {https://ceur-ws.org/Vol-2621/CIRCLE20\_06.pdf},
  bibsource    = {dblp computer science bibliography, https://dblp.org}
}

@article{Bodigutla2021,
  author       = {Praveen Kumar Bodigutla},
  title        = {High Quality Related Search Query Suggestions using Deep Reinforcement
                  Learning},
  journal      = {CoRR},
  volume       = {abs/2108.04452},
  year         = {2021},
  url          = {https://arxiv.org/abs/2108.04452},
  eprinttype    = {arXiv},
  eprint       = {2108.04452},
  bibsource    = {dblp computer science bibliography, https://dblp.org}
}

@misc{aisearch2025,
  title         = {Towards {AI} Search Paradigm},
  year          = {2025},
  eprint        = {2506.17188},
  archivePrefix = {arXiv},
  primaryClass  = {cs.IR},
  note          = {arXiv preprint},
}

@inproceedings{LewisRag2020,
    author = {Lewis, Patrick and Perez, Ethan and Piktus, Aleksandra and Petroni, Fabio and Karpukhin, Vladimir and Goyal, Naman and K\"{u}ttler, Heinrich and Lewis, Mike and Yih, Wen-tau and Rockt\"{a}schel, Tim and Riedel, Sebastian and Kiela, Douwe},
    title = {Retrieval-augmented generation for knowledge-intensive NLP tasks},
    year = {2020},
    isbn = {9781713829546},
    publisher = {Curran Associates Inc.},
    address = {Red Hook, NY, USA},
    booktitle = {Proceedings of the 34th International Conference on Neural Information Processing Systems},
    articleno = {793},
    numpages = {16},
    location = {Vancouver, BC, Canada},
    series = {NeurIPS '20}
}

@article{Kalervo2002ndcg,
    author = {J\"{a}rvelin, Kalervo and Kek\"{a}l\"{a}inen, Jaana},
    title = {Cumulated gain-based evaluation of IR techniques},
    year = {2002},
    issue_date = {October 2002},
    publisher = {Association for Computing Machinery},
    address = {New York, NY, USA},
    volume = {20},
    number = {4},
    issn = {1046-8188},
    url = {https://doi.org/10.1145/582415.582418},
    doi = {10.1145/582415.582418},
    journal = {ACM Trans. Inf. Syst.},
    month = oct,
    pages = {422–446},
    numpages = {25}
}

@misc{verma2023,
      title={Seasonality Based Reranking of E-commerce Autocomplete Using Natural Language Queries}, 
      author={Prateek Verma and Shan Zhong and Xiaoyu Liu and Adithya Rajan},
      year={2023},
      eprint={2308.02055},
      archivePrefix={arXiv},
      primaryClass={cs.IR},
      url={https://arxiv.org/abs/2308.02055}, 
}

@inproceedings{Jiang2018,
    author = {Jiang, Jyun-Yu and Wang, Wei},
    title = {RIN: Reformulation Inference Network for Context-Aware Query Suggestion},
    year = {2018},
    isbn = {9781450360142},
    publisher = {Association for Computing Machinery},
    address = {New York, NY, USA},
    url = {https://doi.org/10.1145/3269206.3271808},
    doi = {10.1145/3269206.3271808},
    booktitle = {Proceedings of the 27th ACM International Conference on Information and Knowledge Management},
    pages = {197–206},
    numpages = {10},
    location = {Torino, Italy},
    series = {CIKM '18}
}

@inproceedings{Christiano2017rlhf,
  author       = {Paul F. Christiano and
                  Jan Leike and
                  Tom B. Brown and
                  Miljan Martic and
                  Shane Legg and
                  Dario Amodei},
  editor       = {Isabelle Guyon and
                  Ulrike von Luxburg and
                  Samy Bengio and
                  Hanna M. Wallach and
                  Rob Fergus and
                  S. V. N. Vishwanathan and
                  Roman Garnett},
  title        = {Deep Reinforcement Learning from Human Preferences},
  booktitle    = {Advances in Neural Information Processing Systems 30: Annual Conference
                  on Neural Information Processing Systems 2017, December 4-9, 2017,
                  Long Beach, CA, {USA}},
  pages        = {4299--4307},
  year         = {2017},
  url          = {https://proceedings.neurips.cc/paper/2017/hash/d5e2c0adad503c91f91df240d0cd4e49-Abstract.html},
}

@article{Schulman2017ppo,
  author       = {John Schulman and
                  Filip Wolski and
                  Prafulla Dhariwal and
                  Alec Radford and
                  Oleg Klimov},
  title        = {Proximal Policy Optimization Algorithms},
  journal      = {CoRR},
  volume       = {abs/1707.06347},
  year         = {2017},
  url          = {http://arxiv.org/abs/1707.06347},
  eprinttype    = {arXiv},
  eprint       = {1707.06347},
}

@article{Ziegler2019,
  author       = {Daniel M. Ziegler and
                  Nisan Stiennon and
                  Jeffrey Wu and
                  Tom B. Brown and
                  Alec Radford and
                  Dario Amodei and
                  Paul F. Christiano and
                  Geoffrey Irving},
  title        = {Fine-Tuning Language Models from Human Preferences},
  journal      = {CoRR},
  volume       = {abs/1909.08593},
  year         = {2019},
  url          = {http://arxiv.org/abs/1909.08593},
  eprinttype    = {arXiv},
  eprint       = {1909.08593},
}

@article{Stiennon2020,
  author       = {Nisan Stiennon and
                  Long Ouyang and
                  Jeff Wu and
                  Daniel M. Ziegler and
                  Ryan Lowe and
                  Chelsea Voss and
                  Alec Radford and
                  Dario Amodei and
                  Paul F. Christiano},
  title        = {Learning to summarize from human feedback},
  journal      = {CoRR},
  volume       = {abs/2009.01325},
  year         = {2020},
  url          = {https://arxiv.org/abs/2009.01325},
  eprinttype    = {arXiv},
  eprint       = {2009.01325}
}

@inproceedings{RafailovSMMEF23dpo,
  author       = {Rafael Rafailov and
                  Archit Sharma and
                  Eric Mitchell and
                  Christopher D. Manning and
                  Stefano Ermon and
                  Chelsea Finn},
  editor       = {Alice Oh and
                  Tristan Naumann and
                  Amir Globerson and
                  Kate Saenko and
                  Moritz Hardt and
                  Sergey Levine},
  title        = {Direct Preference Optimization: Your Language Model is Secretly a
                  Reward Model},
  booktitle    = {Advances in Neural Information Processing Systems 36: Annual Conference
                  on Neural Information Processing Systems 2023, NeurIPS 2023, New Orleans,
                  LA, USA, December 10 - 16, 2023},
  year         = {2023},
  url          = {http://papers.nips.cc/paper\_files/paper/2023/hash/a85b405ed65c6477a4fe8302b5e06ce7-Abstract-Conference.html},
}

@article{Ethayarajh2024,
  author       = {Kawin Ethayarajh and
                  Winnie Xu and
                  Niklas Muennighoff and
                  Dan Jurafsky and
                  Douwe Kiela},
  title        = {{KTO:} Model Alignment as Prospect Theoretic Optimization},
  journal      = {CoRR},
  volume       = {abs/2402.01306},
  year         = {2024},
  url          = {https://doi.org/10.48550/arXiv.2402.01306},
  doi          = {10.48550/ARXIV.2402.01306},
  eprinttype    = {arXiv},
  eprint       = {2402.01306},
  bibsource    = {dblp computer science bibliography, https://dblp.org}
}

\appendix
\section{Prompt Templates for Critique--Revision and Generation}
\label{appendix}

This appendix presents simplified versions of the prompts used to train the \emph{Generator}, \emph{Critic}, and \emph{Reviser} in our agentic critique--revision pipeline. The actual production prompts contain additional implementation details and product-specific wording; here we focus on the structure and key instructions so that practitioners can adapt the approach to their own query auto-completion systems.

\subsection{Generation Prompt (Generator)}
\label{app:gen-prompt}

The Generator prompt conditions on a user prefix, historical query candidates, and retrieved app metadata, and asks an LLM to output a formatted list of query completions. An illustrative template is shown below.

\begin{lstlisting}
[SYSTEM]
You are an expert App Store query suggestion assistant. Given a partial user query
and structured data about historical queries and apps, generate up to 10 diverse,
accurate, and helpful query completions that help users discover relevant apps quickly.

[INPUT]
1. User Prefix:
   {prefix}

2. Query Candidates ({query_candidate_count}):
   - Historical queries issued after this prefix, with frequency, conversion rate,
     click-through rate, and example search results.

3. Apps Metadata ({relevant_app_count}):
   - Titles, categories, and short descriptions for apps that appear in the results.

[GUIDELINES]
- Ground every suggestion ONLY in the provided query candidates and app metadata.
- Suggestions must complete or closely match the user prefix and reflect plausible intents.
- Avoid unsafe, harmful, or policy-violating queries (unless they are exact app titles).
- Avoid near-duplicate suggestions that would lead to almost identical result pages.
- Prefer fewer high-quality suggestions over many weak or ungrounded ones.

[OUTPUT FORMAT]
Return only a list of queries between <answer> and </answer>, one per line:

<answer>
query1
query2
...
</answer>
\end{lstlisting}

\subsection{Critique Prompt (Critic)}
\label{app:critic-prompt}

The Critic receives the original generation prompt (\texttt{Prompt}) and the Generator's JSON-formatted output (\texttt{Response}). It evaluates each suggestion along multiple dimensions and decides whether a revision pass is needed. An illustrative template is:

\begin{lstlisting}
[SYSTEM]
You are an expert reviewer for query suggestions in a mobile app store. Given the
generation prompt and the model's response, assess each suggested query and provide
detailed feedback on how to improve the list.

[INPUT]
1. Prompt:
   - Full generation prompt, including user prefix, query candidates, retrieved apps,
     and instructions.

2. Response:
   - JSON-formatted suggestion list and any associated metadata (e.g., origins).

[EVALUATION DIMENSIONS]
For each suggested query, comment on:
- Relevance: is it relevant to the prefix and supported by engagement signals?
- Prefix matching: does it complete or closely match the prefix?
- Fluency and sense: is it grammatical and plausible as a human query?
- App-store focus: does it help users find apps?
- Safety: does it avoid sexual, violent, or otherwise harmful content?
- Groundedness: can it be traced to query candidates or retrieved apps, and do
  any claimed app origins match the metadata?
- Duplication: is its intent redundant with earlier suggestions?
- Coverage: does the overall list reach the desired length with grounded, diverse queries?

[OUTPUT]
- Provide a short paragraph for each query with issues and concrete improvements.
- End with a final decision flag, for example:

Final decision to revise: YES   (or NO)

The Critic does not output a revised list; it only recommends changes.
\end{lstlisting}

\subsection{Revision Prompt (Reviser)}
\label{app:revise-prompt}

The Reviser takes the same generation prompt, the initial Generator response, and the Critic's assessment, and produces an improved suggestion list. An illustrative template is:

\begin{lstlisting}
[SYSTEM]
You are a query suggestion improver for a mobile app store. Using the original
prompt, the initial suggestions, and a detailed critic review, produce a revised
list of high-quality, grounded, and diverse query completions.

[INPUT]
1. Prompt:
   - Original generation prompt.

2. Initial Response:
   - Initial suggestion list and any metadata.

3. Assessment:
   - Critic's per-suggestion analysis and overall revision recommendations.

[REVISION GUIDELINES]
- Respect all constraints from the original generation prompt (safety, groundedness,
  output format, diversity).
- Follow the Critic's feedback to fix unsafe or ungrounded queries, poor prefix
  matching, and near-duplicates.
- Preserve suggestions the Critic deems high-quality; do not change them unnecessarily.
- If the Critic's final decision is that no revision is needed, you may return
  the initial list unchanged.

[OUTPUT]
1. Briefly describe the main changes (e.g., which suggestions were replaced or
   re-ordered and why).
2. Then output the final suggestion list in the same format as the Generator:

<answer>
query1
query2
...
</answer>
\end{lstlisting}

\end{document}